\documentstyle[12pt,psfig,twoside]{article}
\begin{document}
\hbadness=10000
\pagenumbering{arabic}
\pagestyle{myheadings}
\markboth{J. Letessier, J. Rafelski and A. Tounsi}{Strange Particle
Abundance in QGP}
\title{Strange Particle Abundance in QGP Formed\\ in 200 GeV A Nuclear
Collisions}
\author{$\ $\\
\bf Jean  Letessier$^1$, Johann Rafelski$^{1,2}$ {\rm and} Ahmed
Tounsi$^1$\\ $\ $\\
$^1$ Laboratoire de Physique Th\'eorique et Hautes Energies\thanks{\em
Unit\'e  associ\'ee au CNRS UA 280, \rm\newline
\hspace*{0.5cm}
Postal address: LPTHE~Universit\'e Paris 7, Tour 24, 5\`e
\'et., 2 Place Jussieu, F--75251 Cedex 05.}~, Paris
\\ 
$^2$ Department of Physics, University of Arizona, Tucson, AZ 85721\\
}
\date{}   
\maketitle 
\begin{abstract}
{\noindent
We investigate the relative abundance of strange particles produced in
nuclear collisions at the SPS energies ($\sim$ 9 GeV A in CM frame)
assuming that the central reaction fireball consists of quark-gluon
plasma. We show that the total strangeness yield observed in Sulphur
induced reactions is compatible with this picture.
}\end{abstract}  \vspace*{0.5cm}
\begin{center}
{\it Physics Letters} B 323 (1994) 393--400.
\end{center}
\vfill
{\bf PAR/LPTHE/94--02}\hfill{\bf December 1993}
\eject

{\bf 1. Introduction}:
Abundant strangeness production in relativistic nuclear collisions is
suggestive of the quark-gluon plasma (QGP) formation \cite{Raf82}. But
the relative abundances of strange and multi-strange baryons and 
anti-baryons have been suggested as the better signatures
discriminating against other possible forms of highly excited hadronic
matter. These relatively complex composite particles are more
sensitive to the environment from which they emerge and show
pronounced anomalies in their expected yields \cite{RD87} as compared
to experience based on N--N reactions. However, their systematic
measurement constitutes a formidable experimental challenge and only
today experimental results on  their (relative) abundances and spectra
are being reported \cite{WA85,WA85new,WA85OM,NA35QM,NA3693} for  S--A
(A = S, Ag, W, Pb) collision systems. 
 
It was immediately recognized that this data lend itself naturally to
an interpretation in terms of local thermal equilibrium fireball model
and the resulting thermal parameters are suggestive of a source
consisting  of a deconfined QGP fireball \cite{Raf91}. This hypothesis
was further supported by the observation that details of the produced
particle multiplicity point to a high entropy primordial phase
\cite{Let93}. Such a phase could not be brought into consistence with
the properties of a conventional hadronic gas state made of individual
hadrons. Rejection of confined, hadronic gas (HG) phase became even
more compelling after a more complete analysis of (revised) data was
completed allowing also for  resonance disintegration
\cite{Let94,Let94b}. The formation of the primordial high entropy
phase in turn implies that the relative yield of strange to 
non-strange particles will be diluted in the final hadronic state by
the necessity to generate a high particle multiplicity associate with
the high entropy content of the source. Hence the question ensues if
indeed strangeness multiplicity enhancement is visible should  the
hadronic multiplicity increase, as is expected for a QGP central
fireball. We will consider how the theoretical expectations about the
relative abundance of strange to non-strange particles produced from
the thermal QGP phase compare to the results of nuclear collisions for
S--S, S--Ag \cite{NA35QM} reaction systems and comment on S-W
\cite{WA85new} and S--Pb \cite{NA3693} results. We also shall
extrapolate our findings to the forthcoming Pb--Pb collision
experiments.
 
We find good agreement between experiment and the data, once one
allows for the presence of the strangeness occupancy factor
$0.5<\gamma_{\rm s}<1$ \cite{Raf91}, where the value of $\gamma_{\rm
s}$ is believed to depend on the size and life span of the QGP region
formed in the reaction considered \cite{Raf93}. The yield we find is
consistently higher than one has found in conventional N--N reactions
\cite{GH91}, and we predict a still greater enhancement for the Pb--Pb
case. Thus though the global strangeness enhancement seen at one
collision energy cannot be alone seen as convincing evidence for the
formation of QGP state, we find that the magnitude and systematic
behavior seen at 200 GeV as function of the target mass fits well into
the picture of QGP fireball developed to account for the anomalies
found in (multi-)strange (anti-)baryon abundances.
 
{\bf 2. Reaction dynamics}:
Implicit in the physical picture employed here (see Ref.\,\cite{Let94}
for more details) is the formation of a central and hot matter
fireball in the nuclear collisions at 200 GeV A. This reaction picture
presumes that a fraction of energy and flavor (baryon number) content
of the central rapidity region is able to thermalize, characterized by
the temperature $T$ and the chemical fugacities $\lambda_i$ of the
different conserved quark flavors $i=u,d,s$. This local thermal
equilibrium hypothesis has been inferred from the global evidence on
particle spectra and an extensive analysis \cite{ULI93} which included
diverse distortions of the spectra caused by remaining  collective
longitudinal flow of unstopped matter, and resonance decays. For
example we note that the WA80 \cite{WA80spec} collaboration has
presented data regarding $\pi^0$ and $\eta$ temperatures over a wide
transverse mass range of $0.4<m_\bot<3$ GeV. In the domain
$m_\bot>1.5$ GeV where the corrections arising from disintegrations of
hadronic resonances are small one can infer a source temperature of
about $240\pm10$ MeV, very similar to what is seen for the strange
particles produced at such high transverse mass \cite{WA85new}.
 
Given these high temperatures we can subsume the following: early in
the collision there was a maximum compression point at which the
conversion of energy from relative motion into the thermal energy
content of the central fireball stopped either because the kinetic
pressure of colliding matter has being exceeded by the internal
pressure of the fireball at the kinetic energy of the collision, or
because the smallness of the target/projectile the collision reaction
was ended. Evidence for longitudinal flow in rapidity hadron spectra
points to this latter case in the S--A reactions at 200 GeV A energy.
Following the collision compression and the ensuing rapid
thermalization the matter develops collective rarefaction wave which
absorbs progressively greater fraction of the thermal energy, and the
system cools down till the final state particles freeze-out. They are
propelled by the rarefaction wave and hence their spectral temperature
is blue-shifted by the collective flow velocity. Therefore the
temperature observed in the $m_\bot$ particle spectra is nearly the
same that would be recorded directly from the initially highly
compressed state.
 
Thus whichever the exact dynamical history may be, {\it e.g.\/}
surface evaporation or rarefaction followed by low density freeze-out
or combination of both, the particle spectra carry the information
about the temperature at which the heating due to the dynamical
squeeze of colliding nuclei stops. The experimental evidence derived
from strange particle spectra is that the initial state formed in S--A
collisions at 200 GeV A reaches $T\simeq232\pm3$ MeV \cite{WA85new},
while in  S--S  case it is  $T\simeq 194\pm15$ MeV \cite{NA35QM}. It
is not at all self-evident that these results are consistent with the
nuclear reaction process when QGP equations of state are invoked. The
condition of the initial state formed in the collision is controlled
by the stopping of energy and baryon number in the central fireball.
We introduce coefficients $\eta_E,\eta_B$ which describe the fraction
of energy and baryon number deposited in the thermalized fireball. A
typical value arising from global characteristics of the reaction data
is  $\eta_E, \eta_B=60$\% of the CM available energy and the total
baryon number content of S--Pb reaction system consisting of the tube
of projectile matter overlapping with the tube of matter in the target
\cite{Ott93}. The implicit hypothesis that $\eta_E/\eta_B\simeq1$ has
as important consequence that the thermal state energy per baryon is
equal to the initial energy per baryon in the collision system. We use
the following parameters: 
{\begin{table}[htbp]
\caption{ Conditions in the CERN collision systems.}
\begin{center} 
\begin{tabular}{|c|c|c|c|} \hline
System & Momentum&  $\sqrt{s}_{\rm NN}$ (GeV)& $\gamma_{\rm s}$ \\
\hline
S--S/Ag    & 200 GeV A &    9.7/9.4  &$0.56\pm0.2$\\
S--W/Pb & 200 GeV A &    8.8  &$0.8\pm0.1$\\ 
Pb--Pb  & 157 GeV A &    8.6  &1 (?)\\ 
\hline
\end{tabular} 
\end{center} 
\end{table}}

The value of strangeness occupancy factor $\gamma_{\rm s}$ is not
overly important (we show above our final results) in our later study
of equations of state and the evolution of the compressed state.
Since, to a good approximation,  $\gamma_{\rm s}$ enters as a
multiplicative factor in the strangeness energy density 
(typically 20\% of all)  a reduction of
$\gamma_{\rm s}$ by 50\% has similar influence on $E/B$ as has the
uncertainty related to the condition $\eta_E\simeq\eta_B$. We note
that when computing the collision energy of the Pb--Pb system we have
assumed that the projectile momentum per nucleon will be scaled by the
factor $Z/A$ (and correcting for nuclear binding) down from the
$p$--$p$ 400 GeV/c nominal CERN-SPS value.
 
{\bf 3. Chemical conditions and deconfinement}:
Studying the relative strange particle abundances it is possible to
determine precisely the values of quark fugacities $\lambda_i$. Full
analysis of the S--S \cite{Sol93} and S--W data \cite{Let94} obtained
at 200 GeV A has shown that the strange quark fugacity $\lambda_{\rm
s}\simeq 1$, which is not the case for the lower energy AGS results
obtained with 14 GeV~A projectiles \cite{RD93}. The recently reported 
$\overline{\Omega}/\Omega$ result \cite{WA85OM} has provided another
independent indication that $\lambda_{\rm s}\simeq 1.0\pm0.1$
\cite{Let94}. $\lambda_{\rm s}\sim1$ is natural for a directly
disintegrating QGP phase where the symmetry between the $s$ and $\bar
s$ quarks is reflected naturally in the value of $\lambda_{\rm s}=1$.
Since this value is observed in the final state this suggests that the
hadronization of the QGP phase occurs in particular conditions, given
that  in general the HG phase does not maintain this value of
$\lambda_{\rm s}$. In the HG phase, whatever the equation of state,
$\lambda_{\rm s}=1$ is an exceptional condition. At final baryon
number ({\it viz.\/} $\lambda_{\rm d}$ and $\lambda_{\rm u}\ne 1$) the
strangeness conservation constraint requires that the number of
strange and anti-strange valance quarks bound in final state hadrons
are equal. This is in general incompatible with $\lambda_{\rm s}=1$.  
We now turn to the light quark fugacities: as there is only a slight
asymmetry in the number of $u$ and $d$ quarks in the heavy nuclei used
in experiments it is thus convenient to introduce the quark fugacity
$\lambda_{\rm q}^2=\lambda_{\rm u}\lambda_{\rm d}$ and to confine the
asymmetry between the number of neutrons and protons to the parameter
$\delta\lambda\le0.03$ with $\lambda_{\rm d}/\lambda_{\rm
u}=(1+\delta\lambda)^2$ \cite{Let94}. Relative abundances of (strange)
particles can be used to fix the value of the quark fugacity to
considerable precision. For S--W collisions $\lambda_{\rm
q}=1.49\pm0.05$ is found \cite{Let94}, while for S--S collisions at
central rapidity, $\lambda_{\rm q}\simeq 1.36$ \cite{Sol93} best describes 
the data.
 
It is important to note that the presence of a unique value of
$\lambda_{\rm q}$ determining all particle abundances for each
collision system has considerable implications reaching beyond the
observations regarding establishment of chemical equilibrium. Note
that as the highly compressed hadronic system evolves in time, its
entropy must increase or remain constant --- typically it remains
constant since further entropy production is difficult as model
considerations show. More conveniently, one looks at the specific
entropy per baryon $S/B$ which must nearly remain constant even in
presence of considerable particle evaporation --- such emission
processes are likely to reduce the baryon and entropy content of the
fireball at comparable rate. For physical systems such as is a gas of
interacting hadrons, a time evolution at fixed $S/B$ is not compatible
with a fixed value of $\lambda_{\rm q}$ and, moreover, one should
expect, in models in which particle freeze-out occurs subject to their
respective interaction cross sections, different freeze out values of
$\lambda_{\rm q}$. Thus the final state particle abundances for a
given collision system should be described by values of $\lambda_{\rm
q}$ which vary from particle to particle.
 
The exception to these observations corresponds to a temporal
evolution under equations of state in which aside of the dimensionless
variables $\lambda_{\rm q},\ \lambda_{\rm s}$ no dimensional scale
occurs aside of $T$  --- since the specific entropy $S/B$ is a
dimensionless quantity it can in this case be only a function of
$\lambda_{i}$, the $T$ dependence must cancel. The only known hadronic
physical system which satisfies this condition is the QGP phase, and
in order to give a unique value of $\lambda_{\rm q}$ for all particles
it must be hadronizing directly into decoupled hadronic particles, as
we already argued above. Thus presence of unique values of
$\lambda_i=\{ \lambda_{\rm s},\lambda_{\rm q} \}$ (referred to as
presence of chemical equilibrium) is proving to be a strong, even
though indirect evidence for the occurrence of the deconfined state.
   
{\bf 4. The QGP equations of state}:
We now develop a framework in which we can compute the expected
relative abundance of strange particles emerging at the end of the
nuclear collision in which QGP phase was formed. An essential
ingredient in our model are QGP equations of state. Our assumption is
that the gluons $G$ and light quarks $u,\ d,\ \bar u,\ \bar d$ are
considered in full equilibrium (thermal and abundance) while we shall
allow for suppression of the occupancy of  $s,\ \bar s$ quarks by the
factor $\gamma_{\rm s}$. We will allow for `thermal' masses of all
these particles according to the relation:
\begin{eqnarray}
m_i^2=m^2_0+(cT)^2\,,
\end{eqnarray}
where in  principle we have $c^2\propto\alpha_{\rm s}$, but given the
current uncertainty regarding the value of the coefficient $c$ we 
shall simply explore its consequence in the domain  $c\sim 2$ arising
for $\alpha_s\sim 1$ in the standard formulas \cite{thermal}. We take
$m^0_{\rm q}=5\simeq0$ MeV, $m_{\rm s}^0=160$ MeV, and $m_{\rm
G}^0=0$.
 
It has been found in perturbative thermal QCD \cite{Chin78} which has
been confirmed by more sophisticated lattice gauge numerical
calculations that another, often more significant effect of the
interaction is the reduction of the available degrees of freedom. We
implement the following effective counting of gluon and quark degrees
of freedom: 
\begin{eqnarray}
g_{G}=\hspace{-0.6cm}&&16\to 
     16\left(1- {15\alpha_s\over 4 \pi}\right)\,,\nonumber\\
g_{i-{\rm T}}=\hspace{-0.6cm}&&\ 6\ \to 
     \ 6\left(1-{50\alpha_s\over 21\pi}\right)\,,\\
g_{i-{\rm B}}=\hspace{-0.6cm}&&\ 6\ \to 
     \ 6\left(1-2{\alpha_s\over \pi}\right)\,,\nonumber
\end{eqnarray}
where $i=u,\ d,\ s$ and for quarks two factors are needed: the factor
$g_{i-{\rm T}}$ controls the expression when all chemical potentials
vanish (the $T^4$ term in the partition function for massless quarks)
while $g_{i-{\rm B}}$ is taken as coefficient of the
additional terms which arise in presence of chemical potentials. Note
that we treat all three light quark flavors on the same footing and
that in principle these corrections were established only in the limit
$m_i\ll T$, and hence not in the limit here considered when the
thermal mass exceeds the temperature. We favor for the QCD coupling
$\alpha_s=0.6$ but we have explored the dependence of our results on
variations in $\alpha_s$. Given these interaction effects,
numerical integration of the Bose/Fermi distributions for
quarks/gluons allow us to obtain any physical property of the QGP. We
have first made many studies to assess how the different uncertainties
in the parameters and initial state hypothesis affect our results, and
we believe that our key findings here reported are not significantly
affected. 
 
\begin{figure}[t]
\begin{minipage}[t]{0.475\textwidth}
\vspace{-1.6cm}
\centerline{\hspace{0.2cm}\psfig{figure=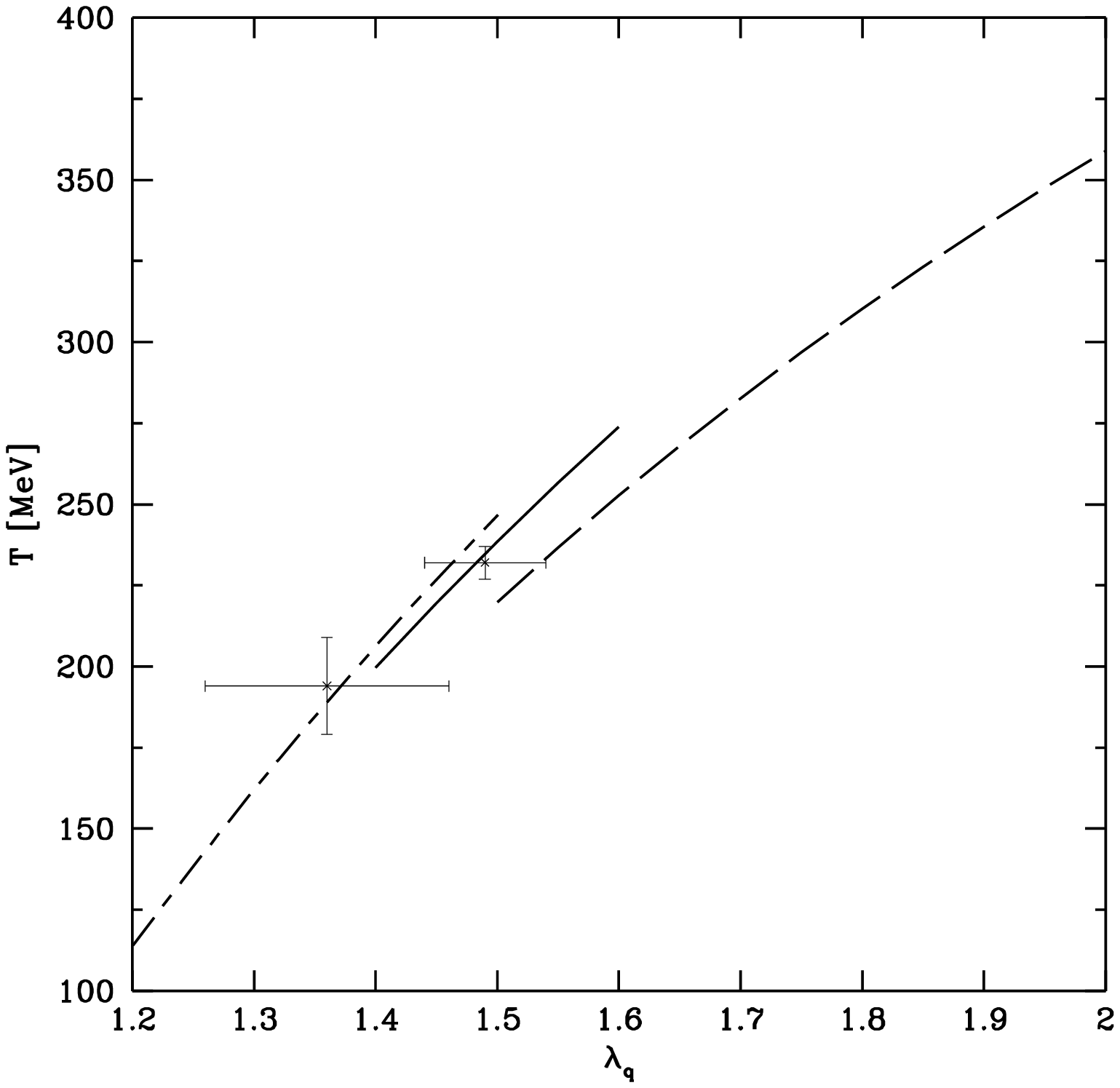,height=11.5cm}}\vspace{-2cm}
\caption[muBT]{\small 
Energy per baryon constraint in the $T$--$\lambda_{\rm q}$ plane for
$\alpha_{\rm s}=0.6$ and $m_i^2=m_0^2+(2T)^2$. Solid line: S--W/Pb for
8.8 GeV A in CM with $\gamma_{\rm s}=0.8$, with the experimental
WA85 point. Long-short dashed curve: S--S for 9.7 GeV A in CM
with $\gamma_{\rm s}=1$, with the experimental NA35 point. The long
dashed line for Pb--Pb for 8.6 GeV A in CM with $\gamma_{\rm s}=1$.
\protect\label{F1}
}
\end{minipage}\hfill
\begin{minipage}[t]{0.475\textwidth}
\vspace{-1.6cm}
\centerline{\hspace{0.2cm}\psfig{figure=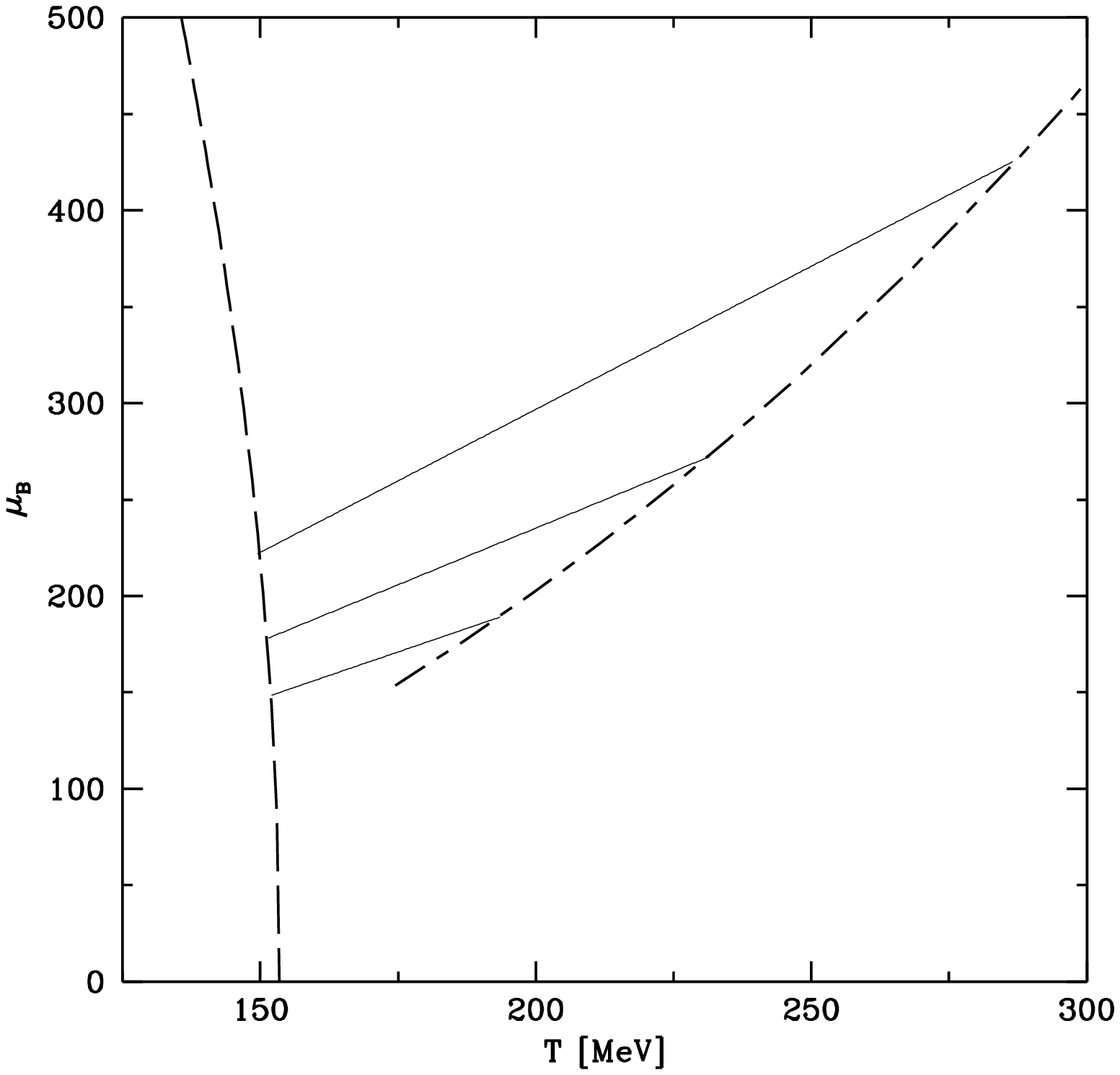,height=11.5cm}}\vspace{-2cm}
\caption[Tlamq]{\small
The lines of fixed $S/B=40,\ 50,\ 60$ (from top to bottom),
corresponding to $\lambda_{\rm q}=1.64,\  1.48,\ 1.38$ in QGP, computed
with same parameters as Fig.\,\ref{F1}, with $\gamma_{\rm s}=0.8$. To
high temperature they end on the line $E/B=8.8$ GeV (short-long dashed
line). Towards low $T$ they end on the hypothetical hadronization
curve near $T=150$ MeV (see Ref.\,\cite{Let94b}). 
\protect\label{F2}
}
\end{minipage}
\end{figure}
{\bf 5. Discussion of results}:
For a given initial state $E/B$, Fig.\,\ref{F1} presents the resulting
constraint between the values of initial temperature $T$ and the quark
fugacity $\lambda_{\rm q}$. The experimental results correspond to the
prior analysis (\cite{Sol93,NA35QM} for S--S, \cite{Let93,WA85new} 
for S--W/Pb). We were stunned to see that for the expected set of the QGP
parameters we have obtained as result an excellent agreement with the
results determined in the prior phenomenological analysis of the
particle spectra regarding $T$ and from relative particle abundance
regarding $\lambda_{\rm q}$ in the initial state. We were thus
encouraged to proceed to study the strangeness yield along the
constant entropy trajectories which are shown for the S--W/Pb case in
the $\mu_{\rm B}$--$T$ plane in Fig.\,\ref{F2}: the short-long dashed
boundary arising for the given $E/B = 8.8$ GeV (assuming equal baryon
and energy stopping, and also $\gamma_{\rm s}=0.8$), the evolution at
$S/B=40$ (initial $\lambda_{\rm q}\simeq1.64$, top solid line), 50
($\lambda_{\rm q}\simeq1.48$, the likely experimental case, middle
solid line), 60 ($\lambda_{\rm q}\simeq1.38$, bottom solid line,
corresponding nearly to the S--S collisions) is as expected from prior
qualitative considerations, along straight lines. They end near 150
MeV, corresponding to the hadronization picture developed previously
\cite{Let94b}.
 
We recall that the entropy content defines the hadronization
multiplicity. To a good approximation one can divide the total entropy
by the entropy per particle in a relativistic ($m\le T$) Boltzmann
gas, $S/N=4$, in order to assess the number of particles emerging in
the hadronization process. Two corrections are to be noted:\\
{\bf i)} a heavy nonrelativistic  hadronic  resonance has a greater
per particle entropy content ($S/N=2.5+m/T+\ldots$),\\
{\bf  ii)} resonance disintegration increases the number of final
state particles.\\
These two effects largely cancel for hadronization at relatively small
temperature and the relationship between the specific entropy and
specific multiplicity remains approximately: 
\begin{eqnarray} \label{SBh}
{S\over B}\simeq4{h\over B}\,,
\end{eqnarray}
(as usual, $h$ denotes the total hadronic multiplicity). Thus the
lines of specific entropy which correspond to fixed $\lambda_{\rm q}$ 
are leading to some determined specific multiplicity in the final
state. Note that the higher $\lambda_{\rm q}$ (greater baryon content)
corresponds to a smaller specific entropy content and thus a smaller
specific multiplicity. This implies at equal initial specific energy
that the initial observable temperature rises with $\lambda_{\rm q}$.
Hence as we move from S--S/Ag to \hbox{S--W/Pb} and finally to Pb--Pb, the
specific multiplicity, and hence $D_{\rm Q}\equiv (h^+-h^-)/(h^++h^-)$
decreases (as usual $h^i$ is the hadron multiplicity of charge $i$).
This systematic behavior may appear surprising, since it implies that
the hotter state has a smaller final per baryon particle multiplicity,
but is completely logical and consistent considering that the initial
condition is determined by specific energy which hardly changes, hence
a baryon density increase implies energy density increase. This
justifies the higher initial temperature which transfers a greater
fraction of initial energy into transverse flow, resulting in smaller
specific multiplicity. We note that in the S--Ag system $\lambda_{\rm
q}$ is near the value 1.38 for S--S and certainly smaller than
$\lambda_{\rm q}=1.48$ for S--W/Pb and hence it is at most 5 units of
specific entropy different from the  S--S system. This difference
translates into a fraction (about 0.4) difference in $h^-$ yield per
participant (reduction for S--Ag as compared to S--S) and is thus
hardly observable: this fact was noted already in the recent NA35
reports: the yield of $h^-/B=1.8\pm0.2$ seen by NA35 \cite{NA35QM},
though 50\% above the yield of HG \cite{Sol93} is the same for the
S--S and S--Ag systems. However the value given is lower than the
expectations based  on the specific entropy content of the QGP phase
discussed here, considering that $S/B\simeq60$. There are a number of
possible explanations of this discrepancy: the baryon content is
overestimated due to substantial and asymmetric contamination by
$K^\pm$ of the central region; the formation of QGP in only a fraction
of events or only in a fraction of the volume; etc.. 
  
This decrease in the specific entropy as we go to baryon richer
systems is at the origin of the here important QGP result: we find
that the ratio of strange quark abundance to the entropy content in
the QGP phase is not $T$ dependent, and moreover it depends little on
the initial state, since the $\lambda_{\rm q}$ (= Const. at fixed
specific entropy) dependence is much slower than the canceling
$T$-dependence. We obtain:
\begin{eqnarray} \label{shth}
s_h\equiv{\langle s+\bar s\rangle^{\rm QGP} \over (S/4)^{\rm QGP}}
=0.25\cdot(1+0.47 (\alpha_s-0.6))\cdot\gamma_{\rm s}\,,
\label{sall}
\end{eqnarray}
in the entire parameter region of interest to us. 
 
We now relate the theoretical value $s_h$ to experimental observables.
By virtue of Eq.(\ref{SBh}) the entropy content is related to the
hadronic multiplicity generated in the interaction, and a suitable
measure of the hadronic multiplicity is the sum of baryon content with
the negative hadron multiplicity $h^-$ multiplied by the factor 3 to
allow for the positive and neutral particles:
\begin{eqnarray}
{S\over 4}\simeq{h}\simeq 3h^-+B\,.
\label{sh}
\end{eqnarray}
Aside of the primordial strangeness there is some contribution arising
from gluon fragmentation \cite{RD87}. Considering that each gluon
hadronizes with 15\% probability \cite{RD87} into a $s\bar s$ pair we have:
\begin{eqnarray}\label{scorr}
\langle s+\bar s\rangle=\langle s+\bar s\rangle^{\rm QGP}
\left(1+0.3 {{\langle G\rangle^{\rm QGP}}\over 
{\langle s+\bar s \rangle^{\rm QGP}}}\right)\,.
\end{eqnarray}
Since $\langle G\rangle^{\rm QGP}/\langle s+\bar s \rangle^{\rm
QGP}=0.52$ is akin to Eq.\,(\ref{shth}) practically independent of the
statistical variables, gluon fragmentation increases the observable
final state strangeness yield by 15\%. This `experimental' yield is
approximately:
$
\langle s +\bar s\rangle \simeq 2 K^0_{s}+K^++K^-+2(\Lambda +\Sigma^0+
\overline{\Lambda}+\overline{\Sigma^0})\,,
$
We note that one can replace $K^++K^-$ by  $2K^0_s$ as long as the
isospin asymmetry is not too large. The factor 2 in front of the
neutral hyperon and anti-hyperon abundance accounts qualitatively for
the usually unobserved charged hyperons $\Sigma^\pm$, which account
for about factor 1.6 and the higher strange particle resonances which
remain unobserved, including hidden strangeness ({\it e.g.\/}
$\eta,\eta'$); this `factor two' rule has proven itself in 
A--A collisions\footnote{This prescription gives about 10\% less
strange particle yield compared to another procedure developed in
\cite{Bial92}.} \cite{Let94}. In view of this discussion we thus
define:
\begin{eqnarray} \label{shexp}
s_h^{\rm exp}\equiv {{\langle s+\bar s\rangle}/1.15\over {3h^-+B}}
\end{eqnarray}
and proceed with our discussion as if $s_h^{\rm exp}\simeq s_h$, see
Eq.\,(\ref{shth}). We note here that the advantage of considering the
total strangeness rather than the enhancement of some individual
fraction, as is often common when discussing enhancements of kaons or
hyperons, is that the sum of strangeness abundance, whatever the model
used, is independent of the chemical composition of the source ({\it
i.e.\/} of $\lambda_{\rm q}$) and hence one can easier compare
different collision systems, in which one expects different values of
$\lambda_{\rm q}$.
 
\begin{figure}[t]
\vspace*{-2cm}
\centerline{\hspace{0.2cm}\psfig{figure=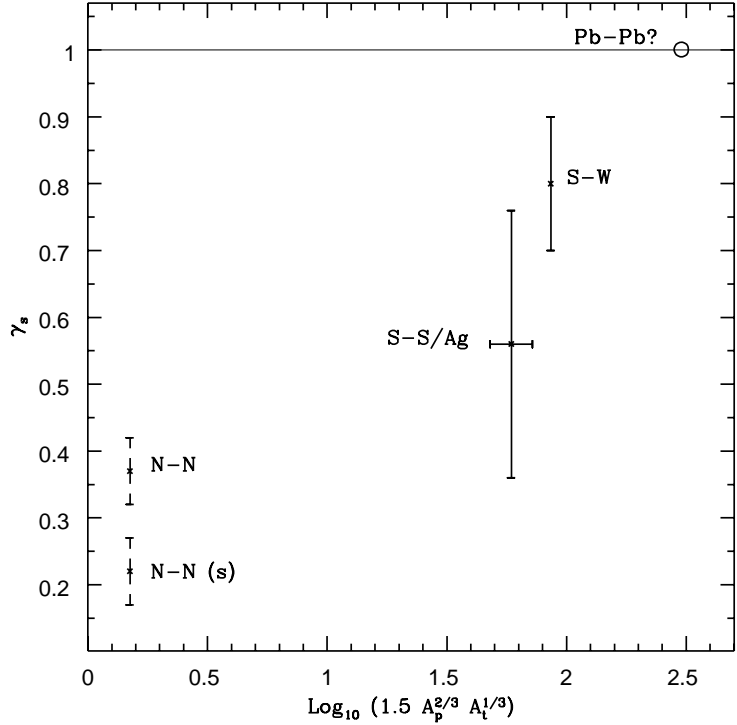,height=14.5cm}}\vspace{-2.4cm}
\caption[gammas] 
{\small
The strangeness occupancy  factor $\gamma_{\rm s}$ as function of
$\log_{\rm 10}(1.5A_{\rm p}^{2/3}A_{\rm t}^{1/3})$,
(projectile (p) and target (t) nuclei). The cross is as determined
here for the collision systems S--S/Ag. We also show prior results
for S--W/Pb \cite{Let94} and the two points for the N--N system which
does not qualify for the thermal description: the result marked `s' is
based on strange particle yields \cite{Sol93}, the other follows from
the (non applicable) prescription here presented.
\protect\label{F3}
}
\end{figure}
First, we recall the `background' value found in N--N reactions
(judiciously weighted \hbox{p--n, p--p} reactions) which follows from the
results given in Table 11 of Ref.\,\cite{GH91}. Using here
Eq.\,(\ref{shexp}), but omitting the $B$ term in the denominator we
find  $s_h^{\rm NN}=0.092\pm 0.014$. It is unwise to interpret this
result in terms of a thermal model, {\it i.e.} to use Eq.\,(\ref{shth}), but
it is important for us to obtain a point to compare with our A--A
results --- we so find $\gamma_{\rm s}^{\rm N\,N}=0.37\pm0.05$ --- to be
compared to the value $\gamma_{\rm s}^{\rm N\,N}=0.22\pm0.05$ which
was obtained using only strange particle yields \cite{Sol93}. Both
these results are shown in Fig.\,\ref{F3}, where we show $\gamma_{\rm
s}$ for the different reaction systems as function of $\log_{10}(1.5
A_{\rm p}^{2/3}A_{\rm t}^{1/3})$, projectile (p) and target (t), which
is an accepted measure of the relative size of the interaction volume.
The smaller N--N value is marked `s' to stress that the hadronic
multiplicity was not directly involved in its determination.
  
M. Ga\'zdzicki \cite{NA35QM} has given detailed information regarding
the multiplicity of strange particles found in S--S/Ag
collisions --- in portions of his analysis he used knowledge of the
phase space distributions in one system to extrapolate the results of
the other. We therefore combine the two sets of data to obtain
$s_h^{\rm S}=0.14$, which may be interpreted in terms of
Eq.\,(\ref{shth}) to imply: $\gamma_{\rm s}^{\rm S}=0.56\pm0.07\pm0.15$\,.
The second error is our estimate of the systematic error made in
equating the here defined experimental value of $s_h^{\rm S}$,
Eq.\,(\ref{shexp}), with the theoretically well defined quantity
$s_h$, Eq.\,(\ref{shth}). The cross in Fig.\,\ref{F3} indicates this
result, alongside with $\gamma_{\rm s}^{\rm S\,W}=0.8\pm0.1$ which
follows from the analysis \cite{Let94} of S--W strange anti-baryon
data \cite{WA85,WA85new}. The increasing trend of the strangeness
occupancy as the size and presumably life-span of the system increases
(see Fig.\,\ref{F3})  and thus its qualitative agreement with the
expected behavior for a system which is near but not at strangeness
abundance equilibrium, is very encouraging --- we note that as
compared to the N--N results we have considerable rise in strangeness
yield  and that a total yield increase by a factor  2.5--5 is
anticipated comparing Pb--Pb with N--N.
 
We note that in Ref.\,\cite{Sol93} the value $\gamma_{\rm s}^{\rm
S\,S}=1.0\pm0.2$ was obtained in a global particle abundance fit. This
is not inconsistent with the current finding, as in this earlier
approach a constraint was imposed that the relative abundance of
hadronizing strange mesons and baryons follows the relative hadronic
gas yield. Since, we have realized \cite{Let94b} that such a
constraint is forcing a hadronization at $T=190$--$200$ MeV, which
indeed was found in Ref.\,\cite{Sol93}. In this condition no attempt
could be made to understand the enhanced (50\%) yield of predicted
negative hadron multiplicity as compared to experiment. It is this
unexplained $h^-$ enhancement which is at the origin of the here
reported smaller value of $\gamma_{\rm s}^{\rm S\,S}$.
 
Very recently, the experiment NA36 has released its final experimental
abundance for strange particles observed in S--Pb reactions
\cite{NA3693}. However, the data does not cover the total strange
particle abundance, but only particles in the interval $0.6
<p_\bot<1.6$ GeV/c are given, and a further cut is made for rapidity
window, with $1.5<y<3$. So restricted in this phase space window
strangeness yield per {\it total} negative hadron multiplicity leads
to $s_h^{\rm cut}\sim0.02$. This result is not inconsistent with the
expectation $s_h^{\rm S\,Pb}=0.8\cdot 0.25=0.20$, but considerable
further effort regarding extrapolation to the unobserved phase space
regions is needed in order to compare theory with experiment.
 
We have shown in this paper that the hypothesis that QGP plasma state
is formed is quantitatively in agreement with the observed strangeness
enhancement in S--A collisions. Our results may not be misconstrued to
constitute a `proof' of the formation of QGP, but should rather be
seen as supportive evidence that the strangeness enhancement as
reported is in agreement with the expectations based on our current
knowledge about QGP form of hadronic matter. We have further found the
interesting result that the observed primordial temperatures and the
corresponding values of quark chemical fugacity are consistent with
the notion of equal baryon number and energy stopping in S--A
collisions, if QGP equations of state are assumed.
 
While previous strangeness enhancement analysis \cite{Bial92} has
focused on the comparison of different strange particle yields with
reference being the N--N collision system, we have established here a
quantitative comparison with the expectations assuming the best
current knowledge of the QGP state. In particular we are able to show
that all experimental results including also the initial temperature
seen at high $m_\bot$ are consistent with the QGP model in which we
allow for a varying degree of strange phase space saturation, which is 
expected to approach unity as the size of the system increases.  We
find an increase in the value of $\gamma_{\rm s}$ (see
Fig.\,\ref{F3}) as we move from S--S/Ag to S--W/Pb systems. Our
finding of an increasing strangeness occupancy factor with increasing
size of the interaction region proves that we can expect to reach full
saturation in the case of the Pb--Pb collisions, for which we hence
expect strange particle to all hadronic multiplicity ratio $s_h^{\rm
Pb\,Pb}=0.25$, which implies that one in four particles produced will
carry strangeness. It is also worthwhile to recall that we expect
somewhat smaller particle multiplicity per participant and a
significantly greater initial temperature in Pb--Pb collisions
compared to S--W/Pb.
  
\vspace{0.5cm}
{\bf Acknowledgement}: We thank Marek Ga\'zdzicki for careful reading
of the manuscript and valuable comments. J. R. acknowledges partial
support by  DOE, grant DE-FG02-92ER40733 and thanks his co-authors for
their kind  hospitality in Paris.
 

\end{document}